\documentclass{bmcart}

\usepackage[utf8]{inputenc} 
\usepackage[sort]{natbib}
\usepackage{color}
\usepackage{url}  
\usepackage{graphicx}  
\usepackage{subcaption}
\usepackage{amsmath}
\usepackage{amssymb}
\usepackage{pifont}
\usepackage{color}
\usepackage{relsize}
\usepackage{enumitem}
\usepackage{float}
\usepackage[margin=1cm]{caption}
\usepackage{adjustbox}
\usepackage{multirow}

\startlocaldefs
\endlocaldefs

\begin{document}

\begin{frontmatter}

\begin{fmbox}
\dochead{Research}


\title{Estimating Community Feedback Effect on Topic Choice in Social Media with Predictive Modeling}


\author[
   addressref={aff1},                   
   noteref={n1},                        
   email={didelani@lsv.uni-saarland.de}   
]{\inits{DI}\fnm{David Ifeoluwa} \snm{Adelani}}
\author[
   addressref={aff2-1,aff2-2,aff2-3},
   noteref={n1},                        
   email={r-koba@nii.ac.jp}
]{\inits{R}\fnm{Ryota} \snm{Kobayashi}}
\author[
   addressref={aff3},
   email={iweber@hbku.edu.qa}
]
{\inits{I}\fnm{Ingmar} \snm{Weber}}
\author[
   addressref={aff4},
   corref={aff4},                       
   email={grabowicz@cs.umass.edu}
]
{\inits{PA}\fnm{Przemyslaw A.} \snm{Grabowicz}}


\address[id=aff1]{
  \orgname{Spoken Language Systems (LSV) Group}, 
  \street{Saarland Informatics Campus, Saarland University},                     %
  \city{Saarbru\"cken},                              
  \cny{Germany}                                    
}
\address[id=aff2-1]{%
  \orgname{The University of Tokyo},
  \city{Chiba},
  \cny{Japan}
}
\address[id=aff2-2]{%
  \orgname{National Institute of Informatics},
  \city{Tokyo},
  \cny{Japan}
}
\address[id=aff2-3]{%
  \orgname{JST, PRESTO},
  \city{Saitama},
  \cny{Japan}
}

\address[id=aff3]{%
  \orgname{Qatar Computing Research Institute},
  \street{HBKU},
  \city{Doha},
  \cny{Qatar}
}

\address[id=aff4]{%
  \orgname{University of Massachusetts},
  \city{Massachusetts},
  \cny{USA}
}


\begin{artnotes}
\note[id=n1]{Equal contributor} 
\end{artnotes}

\end{fmbox}


\begin{abstractbox}

\begin{abstract} 

Social media users post content on various topics. A defining feature of social media is that other users can provide feedback ---called community feedback--- to their content in the form of comments, replies, and retweets. We hypothesize that the amount of received feedback influences the choice of topics on which a social media user posts. However, it is challenging to test this hypothesis as user heterogeneity and external confounders complicate measuring the feedback effect. Here, we investigate this hypothesis with a predictive approach based on an interpretable model of an author's decision to continue the topic of their previous post. We explore the confounding factors, including author's topic preferences and unobserved external factors such as news and social events, by optimizing the predictive accuracy. This approach enables us to identify which users are susceptible to community feedback. Overall, we find that 33\% and 14\% of active users in Reddit and Twitter, respectively, are influenced by community feedback. The model suggests that this feedback alters the probability of topic  continuation up to 14\%, depending on the user and the amount of feedback. 

\end{abstract}




\end{abstractbox}
%

\end{frontmatter}




\section{Introduction}

Social media allow users to post their own content and receive feedback from their audience. 
Online platforms offer various forms of community feedback, including retweets, comments, replies, up-votes, and down-votes. 
Social impact theory suggests that a large amount of positive social feedback, such as support from friends, encourages individuals to continue the behavior that triggered the feedback~\cite{latane}. 
Social impact theory has been tested by social psychological experiments~\cite{Latane1996,Lee2011}, and there is supporting evidence that social feedback affects consumers' behavior~\cite{Naylor2012,Perez2016}.
These findings are consistent with the seminal operant conditioning experiments showing that animal behaviors are reinforced by rewards~\cite{Skinner1938}.
The concern that operant conditioning affects social media users has been raised recently~\cite{Andreassen2015,Deibert2019}.
However, the choice of topic to post is a higher-level cognitive task, in contrast to lower-level behaviors related to survival~\cite{Maslow1943}, that has not been studied yet. Thus it is not clear whether the community feedback affects the topic choice or not.



Here we investigate the question of how the community feedback affects a user's choice of topic to post on social media. This question has practical implications for the design of social media systems. For instance, recommender systems are often designed to optimize for community feedback and engagement. If the feedback affects topics posted by users, then such recommendation algorithms may inadvertently contribute to the growth of polarizing or biased topics that receive more attention than impartial topics~\cite{Zafar2016}. 
Furthermore, the topics discussed on social media and their evolution are important in modern studies of agenda setting~\cite{McCombs1972,Conway2015}.

The following two challenges hinder us from directly measuring the community feedback effect. 
First, social media users are highly heterogeneous: their profiles range from journalists and organizations with pre-scheduled posting agendas to individuals with private accounts having organic agendas. 
Distinct kinds of users are likely to process community feedback in very different ways and not all will be influenced by it. 
To address this heterogeneity issue, we need enough data about the behaviors of individual users over a period of time. 
The second difficulty lies in measuring and controlling for the external factors that can affect users. 
For example, users post burst of tweets due to events such as golf tournaments and movie releases~\cite{Lehmann2012} as well as debates between political candidates~\cite{Trilling2015}. 
In this situation, topic changes might be incorrectly attributed to social feedback, rather than the external events that truly influenced users to switch their posts' topic. 


Community feedback received considerable research attention in the following two areas. 
First, previous studies showed that positive social feedback to online content tends to improve consumer’s opinion about that content~\cite{muchniketal13science,Grabowicz2015} and increase their willingness to disseminate it to friends~\cite{Grabowicz2015}. 
Our study hypothesizes that this community feedback impacts not only the perception of content consumers, but also the author's choice to post related pieces of content in the future. 
Second, analyses of large online communities suggest that positive feedback increases both user retention and the quality of their future posts~\cite{Cheng2014b}, as well as their activity~\cite{ecklesetal16pnas,Cunha2017}.
Complementing these studies, this research is the first work, to the best of our knowledge, to examine whether community feedback influences the topic choices of social media users.

\subsection{Present work:} 



\begin{figure}[t]
\centering
\includegraphics[width=0.43\textwidth]{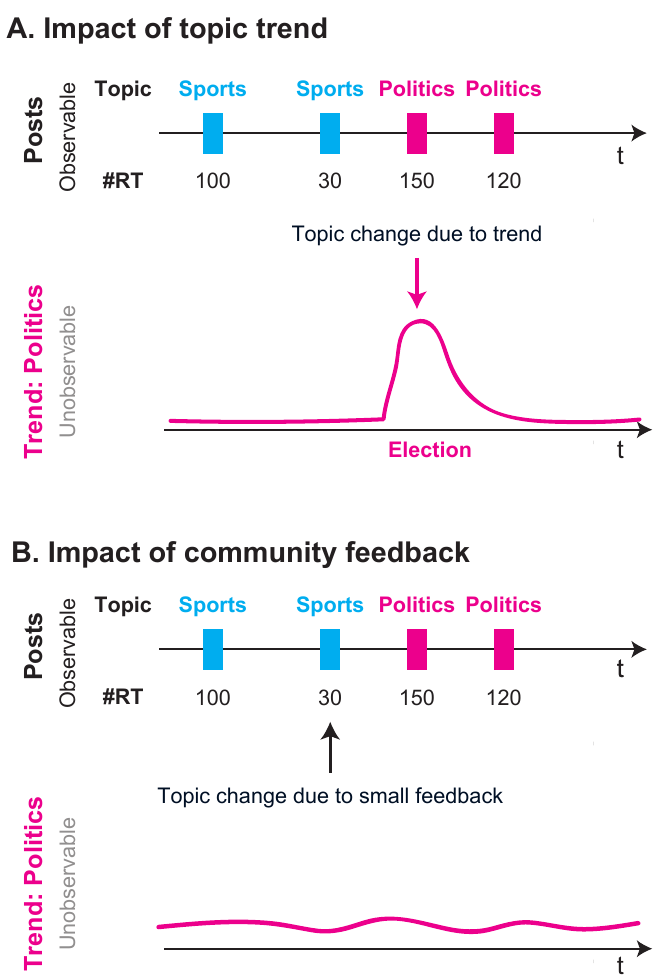}
\caption{Topic trend might be confounders for measuring the community feedback effect on individual users. 
Activities of two users are depicted: User A changed the post topic due to the global topic trend caused by an external factor (Election); 
User B changed the post topic due to feedback (Decrease in the number of retweets). 
This demonstrates the need to control for the topic trend as confounders. In this study, we develop a predictive model that distinguishes the topic trend from the feedback effect. 
}
\label{fig:Schem}
\end{figure}

We hypothesize that the amount of community feedback influences an author's decision to change or continue a topic in their consecutive posts on social media. 
To examine this hypothesis, we develop a semiparametric model of topic continuation and explore which factors influence the probability of topic continuation (Fig.~\ref{fig:Schem}). 
In this model, we incorporate an unobservable confounding factor --- the global topic trend --- that can potentially affect the estimate of the community feedback effect. 
Model-based studies are vulnerable to model misspecification, which may lead to a biased estimate. To address this potential issue, we draw inspiration from philosophy of science by seeking to discover model structure based on its predictive accuracy~\cite{Forster1994}. To this end, we model the topic trend as a flexible time series and learn it directly from data. To further diminish the risk of model misspecification and examine how other factors affect topic choice, we test various structures of the model components, i.e., community feedback and author properties, by optimizing the prediction accuracy.

In this way, we identify two essential factors for topic change --- author's topic preferences and global topic trends --- and demonstrate that our model achieves high predictive accuracy (82\%) for datasets from two social media platforms (Reddit and Twitter). 
We then use this predictive model to quantify how community feedback affects individual users.
While it does not significantly affect most users (67\% for Reddit and 85\% for Twitter), most affected users exhibit a positive rather than negative effect: users tend to continue with the same topic if they receive a significant amount of feedback. 

The contributions of this paper are summarized as follows: we
\begin{itemize}
	\item   develop a predictive model of author's topic continuation,
	\item 	identify key factors for the topic change, including external confounders that are hard to measure, and
	\item   evaluate how community feedback affects individual users in Reddit and Twitter.
\end{itemize}

The remainder of this paper is organized as follows: 
Section 2 surveys the related works, Section 3 describes the datasets from Reddit and Twitter, and Section 4 describes our predictive model for authors' topic changes.  
In Section 5, we examine whether the proposed method can accurately estimate the feedback effect by using synthetic data.
In Section 6, we demonstrate that this model can extract the topic trends and quantify the community feedback effect and individual authors' susceptibility to the feedback. 
In Section 7, we discuss our conclusions and expand on the relation between our results and existing works.

\section{Related Works}

\subsection{Descriptive Studies of Feedback in Social Media}
Whereas our work seeks to quantitatively model the role of social feedback for one particular type of behavior, specifically topic choice, there is rich and insightful body of work that studies user's expectations concerning social feedback \cite{grinbergetal17cscw,frenchbazarova17jcmc,scissorsetal16cscw}. 
This line of work uses surveys to ask users: (i) about their expectations concerning who is likely engage with their content on social media, and (ii) how they feel when these expectations are (not) fulfilled. Researchers report both a positive feeling of connectedness if feedback is received, and a feeling of disappointment when expectations are not met. 
Importantly, not only the feedback quantity but also who provides feedback is part of these expectations and affect the resulting feelings.
Somewhat related is the observational study by Grinberg et al. \cite{Grinberg2016} who examined the activity change before and after posting on Facebook. They hypothesize that the observed increase in activity level after posting may be partly due to  ``anticipation of new interactions''.

Research on the ``imagined audience'' \cite{litt12jbem} expands on the angle of whom a user expects to read and react to a particular social media post. Through surveys, Marwick and Boyd \cite{Marwick2011} studied the techniques used by participants to manage things such as different target audiences on Twitter, with different pieces of content intended to reach different audiences. Empirical work by Bernstein et al. \cite{bernsteinetal13chi} studied the imagined audience by combining large-scale log data on the actual readership and engagement with surveys on a user's expectation. 
They showed that users tend to dramatically underestimate the actual reach of their content and that public signals such as comment or like counts do not strongly indicate audience size.

\subsection{Measuring Social Influence in Social Media}
Concerning the measurement of social influence in social media there are two main approaches.
The first approach is based on experiments, e.g., A/B testing, where users are randomly allocated into treated and untreated groups and the treatment effect is evaluated by comparing the two groups. 
Lab experiments can control some, but not all, external factors (e.g., controlling for global events would require isolating individuals for long time), and face the issue of external validity, compared to field experiments.
Individual effects can be measured if multiple samples are taken from each individual, which may be intrusive to users. 
Thus, well-controlled experiments are not always tractable owing to such practical and ethical limitations. 
Due to these difficulties, few experimental studies have been conducted in social media analysis~\cite{Grabowicz2015,Aral2012,Kramer2014}. 
Two experimental studies that examined the effect of social feedback showed a herding effect of prior positive feedback for social news \cite{muchniketal13science}, and higher peer feedback increases the activity for the receiving user \cite{ecklesetal16pnas}.

The second approach is based on the observational data without any intervention. 
A standard framework for controlling variables is 
the matching methods~\cite{Rubin1974,Stuart2010}, which have been widely used in social media analysis~\cite{Cheng2014b,Cunha2017,Kiciman2018} and has been applied to textual data~\cite{Olteanu2017}. 
%
This approach has two steps: (i) defining the treated and untreated groups from the data, and (ii) controlling the variables that might influence the outcome. 
Although the matching methods are powerful when controlling for multiple external variables, defining the treated and untreated groups in observational data is not always clear. For instance, in our study, the treatment variable is the amount of community feedback, which generally is not binary. 
In such circumstances, more sophisticated propensity score matching can be used to estimate the dose-response relationship~\cite{Imai2004,Hirano2005}.
However, matching without additional modeling components does not account for unobserved confounders, which are present in our context in the form of events happening outside of social media, potentially skewing authored topics to the topics of these events.

In this study, we utilize a regression model with a hidden variable to identify the effect of community feedback on user's choice to continue a topic from observational data. 
Our approach is similar to the structural equation model (SEM) for causal inference~\cite{Shimizu2006,Hoyer2009,Spirtes2016} in the sense that both approaches assume a regression model between the cause and the effect.
There are mainly two differences between SEM and the proposed method. 
First, whereas the SEM aims to discover the causal direction from the data, the proposed method does not identify it. 
We know the direction of the cause and effect, because a potential cause should occur before the effect, and the social feedback was accumulated before the topic choice, so the feedback might cause topic continuation, but not vice versa.
Second, the proposed method is more flexible than traditional SEMs, because we do not specify the functional form of the time series of topical trends. Instead, we learn it directly from data based on a weak specification, which is related to semiparametric approaches to causal inference~\cite{Hoyer2009, Bhattacharya2020}.
Advantages of the proposed method are as follows: (i) it does not require researchers to define the treatment and control group, simplifying the estimation of individual effects, (ii) 
it is straightforward to incorporate a continuous treatment variable,
and (iii) it allows researchers to model unobserved confounders that impact multiple individuals.
However, predictive models may give biased results if they are misspecified. For this reason, various definitions of each model component need to be tested and their structure must be selected based on their predictive accuracy on a hold-out test data or a model selection criterion~\cite{Gelman2014}. 
We will discuss this limitation in detail in Discussion.


\subsection{User Modeling in Social Media} 
In this study, we develop a model of user behavior that predicts an author's topic continuation on social media. 
Related research endeavors pertaining to social media introduced user behavior models predicting 
retweeting behavior~\cite{Yang2010,ZhangQ2015} and post topics~\cite{Xu2012,Yin2014}. 
The user behavior model we develop is similar to the one proposed by \cite{Xu2012} in the sense that both models incorporate the effect of user interests and exogenous factors.
All previous studies have focused on improving either prediction accuracy or recommendation performance. 
Conversely, our investigation uses an interpretable predictive user model to understand the treatment effect of community feedback. 
Furthermore, this is the first study to focus on topic changes in user posts on social media. 
%


The author behavior model developed in this study is motivated by the concepts of endogenous and exogenous factors in the activity on the Web and social media. 
Endogenous factors are defined as interactions within social media or social networks, while exogenous factors are external influences on a community, such as news, catastrophes, and social events, which typically happen externally to social media. 
Previous studies have suggested that these factors can affect the shape of peaks in popularity profiles such as in search queries on Google, viewing activity on YouTube~\cite{Crane2008}, and the adoption of hashtags on Twitter~\cite{Lehmann2012,Fujita2018}. 
We develop a predictive model incorporating topic trends as exogenous factors and a method for estimating these factors from observational social media data. 


\section{Dataset}
\label{sec:data}
We investigate the effect of community feedback on the users' posting behavior based on two popular online discussion platforms: Reddit and Twitter. 
We collected the posts created by thousands of active users: so-called submissions in Reddit and tweets in Twitter. 
Then, for each of these posts, we gathered the community feedback in the form of comments in Reddit and retweets in Twitter. While there are other kinds of feedback available in these platforms, such as up/down-votes or likes (which we study in Appendix A), we focus on these two feedback types, because only for them we are able to collect the time stamps of their creation. Having these timestamps is required for estimating the community feedback at the moment when the author creates their next post, because this feedback can causally influence topic choice.
We gathered this data for the period of six months, between January 1 $-$ June 30, 2016. 
Table~\ref{Tab:data_stat} summarizes the statistics of these datasets. 

\subsection{Reddit}
Reddit~\footnote{\url{https://www.reddit.com}} is a social news and discussion website. 
Users can submit pieces of news or content (e.g., links, videos, and texts) and can vote and comment on these submissions. The submissions are organized by categories, called {\it subreddits}, which cover a variety of topics, ranging from politics and science to sports and entertainment. Each user can subscribe to, post in, and comment in multiple such subreddits.
In our predictive model, we treat each subreddit as a different topic and model the probability that a user continues to post within the same subreddit.


We downloaded submissions and comments from 100 active subreddits using the Reddit data shared by pushshift.io.\footnote{\url{https://files.pushshift.io/reddit/}} 
Previously identified gaps in this data \cite{gaffneymatias18plos} do not apply to our collection because the pushshift.io data had been updated since the gaps were pointed out.
The active subreddits were extracted according to the following procedure. 
After extracting the top 1,000 subreddits with the highest number of subscribers, we selected the top 100 subreddits receiving the most comments and removed inactive users who posted less than 50 posts in the six month. 
In addition, we estimated the fraction of positive, neutral, and negative comments in these subreddits, using a sentiment analysis tool for social media texts \cite{Hutto2014}. We found that positive comments, which may have positive effect on topic continuation, are more common than other comments. 

\begin{table}[t]
\centering
  \begin{tabular}{lll}
    &  Reddit &  Twitter\\     \hline
    Number of posts & 781,614  &  4,437,468\\
    Number of posts with feedback & 696,931 & 4,191,305\\
    Number of users & 7,072 & 6,882\\
    Mean number of posts per user    & 111 & 680  \\
    Mean amount of feedback per user & 30  & 40  \\     \hline    
    \end{tabular}    \vspace{0.3cm}
  \caption{Statistics of the two online communities analyzed in this study.}
  \label{Tab:data_stat}
\end{table}

\subsection{Twitter}
Using Twitter's streaming API, we collected tweets posted by experts and their retweets. 
According to previous works \cite{Ghosh2012,przemyslaw}, an expert is defined as a user satisfying four criteria: (i) they have less than one million followers, (ii) they receive at least 10 retweets per tweet, (iii) they have posted at least 50 tweets during the 6 months, and (iv) they tweet predominantly in English (at least 95\% of tweets in English). 
Most of these users are estimated to be humans (72\%), as opposed to organization accounts, by Humanizr tool \cite{McCorriston2015}.
These expert users are also less likely to be bots, since they tend to be verified accounts that are in multiple Twitter lists~\cite{Ghosh2012}.
We first downloaded the tweets posted by experts, and the retweets were crawled a few months after the tweets were posted. The tweets whose retweets could not be crawled were excluded from our analysis.
Tweets have a limited number of characters, so users who want to publish a longer post are forced to represent it as multiple tweets posted in a quick succession, what is know a tweet thread.
To~remove such threads, we discarded a tweet if the consecutive post is published within less than 30 seconds. 
Then, again, we removed inactive users who posted less than 50 tweets.

For each original tweet authored by experts, we identified its topic, using a topic model. 
We first filtered out the retweets and replies and removed URLs and stop-words from the original posts. Then, we obtained a topic for each tweet by using Twitter-LDA \cite{Zhao2011}, with the number of topics $K$ set to 100, unless stated otherwise.  
Our explorations suggest that our results are qualitatively not affected by the choice of $K$ (Appendix B).

\section{Modeling Author Behavior}
\label{sec:method} 
Our study attempts to quantify to what extent each author is susceptible to the community feedback by using a predictive model. 
Here, we describe the model, its components, and the procedure for fitting model parameters.

\subsection{Model for Predicting Topic Continuation}

We focus on predicting whether an author continues to post on the same topic or not. 
Modeling the phenomenon of topic change, instead of topic selection, reduces the number of samples required to learn model parameters. 
The probability that an author $i$ continues a topic $k$ at time $t$ is described as

\begin{equation}
	P[ Y_i=1 | t, k, f] = S \left( u_i(k; a_i, b)+ g(k,t)+ \alpha_i f \right), 	\label{eq:user_model}
\end{equation}

where $Y_i$ is a binary random variable representing whether the author continues the topic (1) or not (0), $f$ is the community feedback that the user received to their previous post, 
and $S(x)$ is a logistic function. 
We adopted the logistic function, because results of randomized experiments on social influence suggest this parametric family~\cite{Arganda2012,Grabowicz2015} and prior works show that it can be derived as a model of decision-making under social influence, by comparing judgements under uncertainty to Bayesian inference~\cite{Arganda2012}.
In addition, the logistic regression model allows us to incorporate the impact of hidden dynamic confounders, i.e., the topic trends. Importantly, the causal effect is identifiable under this model, because the objective function (Eq. ~\ref{eq:cost}) is concave and model parameters are uniquely identified by its global minimum~\cite{Boyd2004}. 
By contrast, deep neural networks could achieve a better prediction performance, but these models often have multiple local minima~\cite{Choromanska2015}, which results in problems with identifiability. In~addition, the interpretability of such models is limited --- it is a subject of ongoing research and active debate~\cite{Lipton2018}.

Next, we explain each of the three components of this model. 
The first component $u_i(k; a_i, b)$ represents the effect of author properties, where $a_i$ is the user's propensity to continue any topic and $b$ is the effect of the topic preference on the probability of topic continuation.  
We will determine a specific form of user properties by optimizing the model accuracy (Section~\ref{sec:sm}).   
The second component $g(k, t)$ is the effect of the topic trend defined as a flexible time series.
Finally, $\alpha_i$ represents author's susceptibility to the community feedback $f$. 
The feedback $f$ is a function of the number of comments or up/down-votes (Reddit), retweets or likes (Twitter) to author's previous post. Again, we will determine a specific form of the feedback function by optimizing the model accuracy (Section~\ref{sec:sm}).

The central assumption allowing the proposed model to infer the multidimensional missing confounder, i.e., the topic trends, is that all users are affected by this confounder in the same way. 
We show that our model accurately infers such missing confounders by using a synthetic data (Section~\ref{sec:synthtic}).

\subsection{Parameter Inference}
We describe the procedure to specify the model structure and to estimate the parameters. 
For tractability, the topic trend $g(k, t)$ is approximated by a step function with one time interval per day.\footnote{We explored other time intervals and found qualitatively the same results.} This step function is represented by the vectors $\vec{g}_k= \{ g_{k, 1}, g_{k, 2}, \cdots , g_{k, M} \}$, where $k$ represents a topic, and $M= 182$ is the number of time intervals, i.e.\ days in our data. 
In addition, the trend is assumed to be a smooth function of time $t$, which is enforced by L2 regularization~\cite{Kitagawa1996,Kobayashi2011,Kobayashi2019}.

Overall, we estimate $KM+ 2N+ 1$ parameters (approximately 32,300 and 32,000 parameters for Reddit and Twitter, respectively): $\{ \vec{g}_1, \cdots , \vec{g}_K, \vec{a}, b,  \vec{\alpha} \}$, where $K$ is the number of topics, and $N$ is the number of users, 
$\vec{a} = (a_1, \cdots , a_N)$ is a vector of authors' propensities, and $\vec{\alpha}= (\alpha_1, \cdots , \alpha_N)$ is the susceptibility of each author to the feedback. 
The parameters are estimated by maximizing the log likelihood with regularization: 
\begin{eqnarray}
	E( \vec{\theta} ) \! &=& \! \sum_{i, j} \log \left( P[ Y_i= y_{i, j} ] \right)-\beta_u \left\{ \sum_{i=1}^{N} |a_i| + |b| \right\} \nonumber \\
	&& \! - \beta_g \sum_{j=1}^{M-1} \sum_{k=1}^K  (g_{k, j+1}- g_{k, j})^2, \quad \quad 
	\label{eq:cost}
\end{eqnarray}	
where $y_{i, j}$ represents whether the author continues the topic (1) or not (0) in the $j$-th sample, 
and $\beta_{u}$ and $\beta_{g}$ are hyper-parameters controlling the strength of L1 regularization of $a_i$ and $b$ \cite{Bishop2006}, and L2 regularization of $\vec{g}_k$, respectively.
The first term is the log-likelihood for all users, and $j$ is the subscript for respective time window. We learn the hyper-parameters $\beta_u= 0.1$ and $\beta_g= 10$ via a grid search. 
The source code for parameter estimation will be available on GitHub.

\subsection{Evaluation of Prediction Accuracy}
The prediction accuracy is evaluated on a hold-out set of samples, i.e., the last three posts for each user. The training data is resampled from the remaining data using bootstrapping. 
The mean and confidence intervals of accuracy are calculated by repeating this procedure $200$ times, each time using a different training dataset.


\begin{figure}[h]
\centering
  \includegraphics[width=8cm]{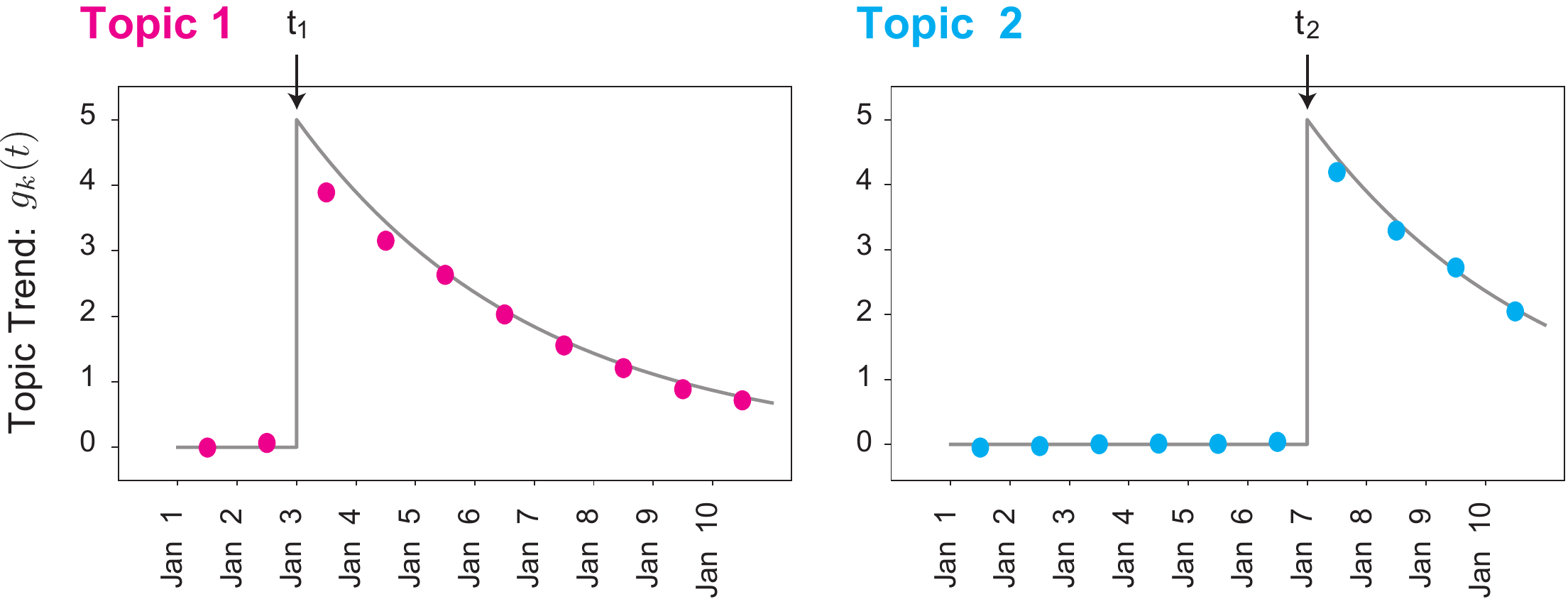}
\caption{Estimating topic trends from synthetic data. Gray lines show the ground truth and the magenta and cyan circles show our estimate. Times of unobserved events are indicated by arrows.}
\label{fig:SynTwoTopics}
\end{figure}

\section{Validation on Synthetic Data}
\label{sec:synthtic}

We examine whether the proposed method (Section~\ref{sec:method}) is able to infer the true susceptibility and topic trend by using a synthetic data.

The synthetic data was generated as follows. 
We first extracted active users who posted more than $50$ tweets in the period of $10$ days (from 1st, Jan., 2016 to 10th, Jan., 2016) together with the timestamps of their posts, which resulted in $1,145$ users. 
Second, we assigned a topic to each post based on the topic continuation model (Eq.~\ref{eq:user_model}). 
We assumed that the number of topics is two, $K=2$, and the first post of each user is assigned as a random topic with the probability 0.5.  
Topics of the subsequent posts are determined by the previous topic and the probability of topic continuation (Eq. \ref{eq:user_model}). 
Third, we specify the model structure, i.e., the user property, the topic trend, and the feedback. 
For simplicity, we assume that the user property is zero, $u_i(k; a_i, b)= 0$, and there are two unobserved events at time $t_1=$ 3rd, Jan. and $t_2= $ 7th, Jan., which impact on the topic trend (Fig.~\ref{fig:SynTwoTopics}). 
The effect of an event on the trends are described as the exponential function, that is, $g_k(t)= g_0 e^{-(t-t_k)/\tau}$ for $t>t_k$ and $g_k(t)=0$ for $t\leq t_k$ ($k=1, 2$), where $g_0=5$ is the amplitude and  $\tau=4$ [days] is the time scale of the user's attention. 
The feedback $f$ is given by a linear function of the trend: $f= c g_k(t)+ \xi$, where $k$ is the topic of the previous post, and $\xi$ is an independent Gaussian variable of zero mean and unit standard deviation. 
The coefficient $c$ controls the correlation between the feedback and the topic trend. If $c > 0$ ($c < 0$), then the feedback is positively (negatively) correlated with the trend.  


Susceptible users are identified based on the proposed model (Eq. \ref{eq:user_model}) and the naive logistic regression, i.e., the proposed model without $g(k,t)$. 
First, we estimate the susceptibility $\alpha_i$ of each user for 200 times by repeating the cross-validation procedure. 
Second, the user susceptibility is determined to be positive (negative) when the lower (higher) bound of 99\% confidence interval of the susceptibility is higher (lower) than zero. The susceptibility is considered to be insignificant if the confidence interval includes zero. 
The model performance is evaluated by the ability to detect susceptible users. All users were randomly divided into two equal groups. 
The susceptibility $\alpha_i$ of a user is set to 1 when they belong to a group, and it is set to 0 when they belong to the other group. 


We examine two cases: (i) the topic trend is a confounder: $c= 1$, and (ii) it is not a confounder: $c= 0$. 
The naive logistic regression fails to detect the susceptible users if the topic trend is a confounder (Table \ref{Tab:Synthetic_A1}), and most users (99 \%) are estimated as significantly susceptible.  
In contract, the proposed method can correctly identify the susceptible users for both cases (Tables \ref{Tab:Synthetic_A1} and \ref{Tab:Synthetic_A0}).
Finally, we show that the proposed model identifies the unobserved confounding events for both topics (Fig.~\ref{fig:SynTwoTopics}), which is the reason why it correctly identifies the susceptible users.
\vspace{0.3cm}

\begin{table}[h]
\centering
\footnotesize
\begin{tabular}{llll}
Group           &  Proposed         &   Baseline      &  True  \\ \hline
\#Positive      &     591(52\%)     &   1,138 (99\%)   &    572 (50\%)  \\
\#Negative      &      34 (3\%)      &      0 (0\%)    &      0 (0\%)  \\
\#Insignificant &    520 (45\%)     &     7 (1\%)    &    573 (50\%)  \\ \hline
Accuracy        &    {\bf 93.5\%}   &     50.6\%      &   -- \\
\hline
\end{tabular}    \vspace{0.3cm}
\caption{
Accuracy of detecting susceptible users when the topic trend is a confounding factor ($c= 1$). ``Proposed'' and ``Baseline'' represent the introduced model with or without the topic trend $g(k, t)$, respectively. Bold letters indicate that the accuracy is higher than 90\%.
}
\label{Tab:Synthetic_A1}
\end{table}

\begin{table}[h]
\centering
\footnotesize
\begin{tabular}{llll}
Group           &  Proposed          &   Baseline           &  True  \\ \hline
\#Positive      &    593 (52\%)      &    537 (47\%)        &    572 (50\%)  \\
\#Negative      &      42 (4\%)       &     24 (2\%)        &      0 (0\%)   \\
\#Insignificant &    510 (44\%)      &    584 (51\%)        &    573 (50\%)  \\ \hline
Accuracy        &    {\bf 92.5 \%}   &    {\bf  91.4 \%}    &     --         \\ \hline
\end{tabular}    \vspace{0.3cm}
\caption{Accuracy of detecting susceptible users when the topic trend is not a confounding factor ($c= 0$). ``Proposed'' and ``Baseline'' represent the introduced model with or without the topic trend $g(k, t)$, respectively. }
\label{Tab:Synthetic_A0}
\end{table}

\section{Results from Social Media Data}
\label{sec:sm}
We investigate the community feedback effect on the authors on Reddit and Twitter. 
The probability of topic continuation calculated from all the posts is 68 \% on Reddit and 46 \% on Twitter.
We develop the predictive model for the topic continuation, i.e., we determine the structure and values of the author properties $u_i(k; a_i, b)$, topic trend $g(k,t)$, and community feedback, $f$, by optimizing the predictive accuracy in a cross-validation setting. 
Finally, we exploit this predictive model to evaluate the community feedback effect on individual authors.  
We investigate author's susceptibility to the feedback, and the effect of the feedback on the prediction accuracy and the probability of topic continuation.

\subsection{Modeling Author Properties and Topic Trend}

\begin{table}[t]
\centering
\footnotesize
   \begin{tabular}{lll}
    Feature &  Reddit &  Twitter \\ \hline
    None                 &  $63.43 \pm 0.00$\%   &  $65.44 \pm 0.00$\%   \\
    Prop                 &  $63.87 \pm 0.00$\%   &  $71.54 \pm 0.01$\%   \\ 
    Pref                 &  $79.35 \pm 0.00$\%   &  $80.29 \pm 0.00$\%  \\
    Prop, Pref           &  $79.30 \pm 0.00$\%   &  $80.52 \pm 0.00$\%  \\
    Prop, Pref, Trend    & $82.27 \pm 0.04\%$    &  $81.97 \pm 0.02$\%  \\
    \hline
    \end{tabular} \vspace{0.3cm}
       \caption{Effect of author property and topic trend on the prediction accuracy. 
       We examined three features: the user-dependent propensity to continue any topic (``Prop''), the preference to topics (``Pref''), and the topic trend due to news and social events (``Trend''). 
       The mean and the 99\% confidence interval are shown. 
       }
    \label{Tab:Pred_Accuracy}
\end{table}

\begin{figure*}[!t]
 \centering
  \includegraphics[width=12cm]{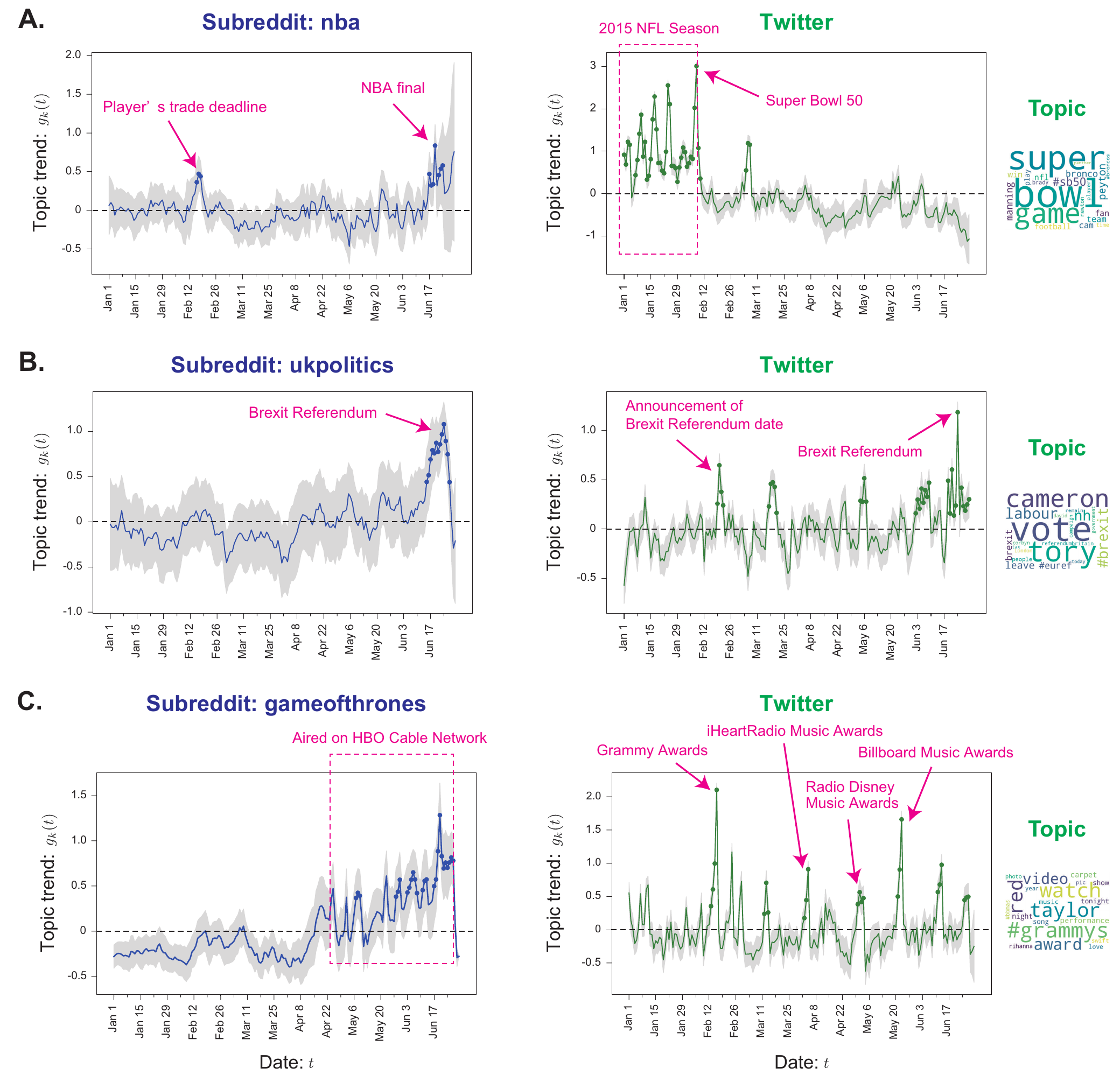}
	\caption{Our predictive model can extract the topic trends which are related to news and social events. 
	Topic trend extracted from three topic categories (A. Sports, B. Politics, and C. Entertainment) are shown. 
	The blue and green lines represent the mean topic trend of Reddit and Twitter, respectively, and the gray area represents the 95\% confidence interval. Filled circles mark the days when the topic trend estimate is significantly higher than zero for at least three consecutive days. Social events manually identified from Wikipedia are written in magenta.}
    \label{Fig:Topic_Trend}
\end{figure*}

We consider two kinds of author $i$'s properties: 
(i) the propensity to continue any topic, $a_i$, and (ii) the effect of the topic preference, $b$. 
These properties are included in the model as $u_i(k; a_i, b)= a_i + b x_i(k)$, where $x_i(k)$ represents the preference of a topic $k$ by a user $i$. 
The propensity captures a tendency of a user to repeat any topic, e.g., users posting in bursts are more likely to continue a topic, because of their bursty activity.
The topic preference captures the bias in posted topics by a user. 
The probability of posting a topic $k$ for a user $i$ was estimated by add-one Laplace smoothing~\cite{Russell2010}: 
$P_i(k) = \frac{N_i(k)+1}{N_i+K}$, where $N_i(k)$ and $N_i$ are the number of posts on topic $k$ and that of all the posts by the user, respectively. 
The topic preference is included in the model as $x_i(k) = S^{-1}(P_i(k) )$, to ensure that the probability of repeating a topic is equal to the null model of posting based on the topic probability $P_i(k)$, if the other factors are not present, i.e., $a_i=0$, $b=1$, $g_{k,j}=0$, and $\alpha_i=0$.
Next, we examine the performance of the models with the various features in predicting whether an author continues to post on the same topic or not (Table~\ref{Tab:Pred_Accuracy}). 
Here, we evaluate the predictive performance using accuracy, i.e., the fraction of correct predictions among all predictions measured on the test set. 
The topic preference, $b$, largely increases the predictive accuracy by 16\% and 15\% for Reddit and Twitter, respectively. 
The propensity to topic continuation, $a_i$, is less important feature than the topic preference, which increases the accuracy by 0.4\% and 6\% for Reddit and Twitter, respectively.
We tested other definitions of both $P_i(k)$ and $x_i(k)$, but this definition resulted in the best predictive accuracy. 
Overall, these author properties explain $80\%$ of authors' decisions to continue a topic.
%
%
The inclusion of topic trend into the model significantly increases the accuracy by $3.0\%$ and $1.5\%$ for Reddit and Twitter, respectively (t-test: $p < 10^{-20}$). 
The effect size (Cohen's $d$) was 17.9 and 17.1 for Reddit and Twitter, respectively, which indicates that the effect of the topic trend is huge~\cite{Cohen2013}.
In addition to this, we evaluate the prediction performance using the F1 score and the Matthews correlation coefficient~\cite{Matthews1975,Kobayashi2019} (Appendix~C). The result is qualitatively the same as that based on the accuracy (compare Table~\ref{Tab:Pred_Full_Reddit} and~\ref{Tab:Pred_Full_Reddit} in Appendix~C with Table~\ref{Tab:Pred_Accuracy}).

%
Figure~\ref{Fig:Topic_Trend} shows three examples of the topic trend extracted from Reddit and Twitter dataset. 
We define the significant period of topic trend as a period of at least three days in which the topic trend is significantly larger than zero.
Interestingly, most of the peaks can be interpreted as popular events and news:
\begin{itemize}
  \item {\bf Sports (Fig.~\ref{Fig:Topic_Trend}A)}: 
        The trend for the subreddit about ``nba" increases around the time of NBA players' trade deadline and the NBA Finals. 
        The Twitter topic trend related to the NFL increases during the NFL season and the Super Bowl championship game. 
  \item {\bf Politics (Fig.~\ref{Fig:Topic_Trend}B)}: 
        The trend for ``ukpolitics" subreddit and the corresponding trend of Twitter topic about politics in the UK exhibit a large peak around the time of the Brexit referendum. 
  \item {\bf Entertainment (Fig.~\ref{Fig:Topic_Trend}C)}: 
        The trend for ``gameofthrones" subreddit (an American fantasy drama television series) dramatically increases during the period when the drama was aired on HBO cable network. The trend of Twitter topic about music and entertainment exhibit peaks before and after famous music awards, e.g, Grammy Awards. 
\end{itemize}
While most of the peaks in the topic trend are interpretable, there exist multiple peaks that we were unable to interpret (e.g., three middle peaks in Fig.~\ref{Fig:Topic_Trend}B for Twitter), either because we lack the knowledge to interpret them or they correspond to the effects present only in some specific groups of users.
Finally, we note that the estimated topic trend differs from the time series of the topic popularity, although the two are correlated (Appendix D). The differences have two origins: (i) the topic trend incorporates, in addition to its popularity, the information about topic's impact on user's decision to continue posting about the topic and (ii) the topic trend is smoother by design, to maintain model compactness and avoid overfitting.

\subsection{Quantifying Community Feedback}
%

We consider the community feedback as a function of the amount of feedback $n_i$, i.e., the number of comments in Reddit and retweets in Twitter that a user $i$ receives for their previous post. 
In addition, we also evaluate the feedback based on up/down-votes in Reddit and likes in Twitter (Appendix A).  
We take into account the following two observations to define the feedback function.
First, the amount of feedback depends on the duration $\Delta t$ from the previous post. The longer the duration is, the more feedback the author will receive.  
Second, previous works showed that a user's feeling is associated with the feedback amount relative to their expectations~\cite{grinbergetal17cscw,frenchbazarova17jcmc,scissorsetal16cscw}.  

We consider three functions of community feedback: (i) the feedback rate, $r_i= n_i/\Delta t$, where $n_i$ is the feedback amount and $\Delta t$ is the duration from the previous post, 
(ii) the logarithm of the feedback rate, $\log(r_i)$, and (iii) the cumulative probability of the feedback rate, $P( R_i< r_i)$, 
where $P(R_i <r_i)$ is the probability of receiving feedback smaller than $r_i$ for the $i$-th user.
We also consider the feedback functions based on the feedback amount $n_i$, in place of the feedback rate $r_i$; however, it does not improve the predictive performance (Appendix E).
In the remainder, we adopt the cumulative probability, $P( R_i< r_i)$, as the feedback function, because it achieved the best prediction accuracy among these candidates. 
The addition of community feedback to the model improves the predictive accuracy by 0.14\% for Reddit and 0.1\% Twitter (Tables~\ref{Tab:Pred_FAmount} and~\ref{Tab:Pred_FRate} in Appendix E), so this effect is many times smaller in comparison to the effect of author properties and topic trend.
Note that this feedback function is a percentile computed with respect to all posts of a user, so it takes into account user's expectations.


\subsection{Susceptibility to Community Feedback}

Next, we evaluate the effect of community feedback on topic continuation separately for each author, by analysing the susceptibility $\alpha_i$ (Eq. \ref{eq:user_model}).
Same as for the synthetic data analysis (Section ~\ref{sec:synthtic}), 
the susceptibility of a user is considered to be positive (negative) when the lower (higher) bound of 99\% confidence interval of the susceptibility $\alpha_i$ is higher (lower) than zero. 
%
\begin{table}[t]
\centering
    \begin{tabular}{lll}
    Group       &  Reddit       &  Twitter  \\ \hline
    \#Positive          &   2,327 (32.9\%)  &    946 (14\%)    \\
    \#Negative          &    13 (0.2\%)    &     85 (1\%)     \\
    \#Insignificant     &   4,732 (66.9\%)    &    5,851 (85\%)    \\
    \hline
    \end{tabular}
         \caption{Distribution of the susceptibility $\alpha_i$ to the community feedback. Positive and negative mean that the susceptibility is significantly higher (lower) than 0, based on $99\%$ confidence interval.}
    \label{Tab:PosNeg_Authors}
\end{table}
%
%
Table \ref{Tab:PosNeg_Authors} shows the distribution of users with positive and negative susceptibility in Reddit and Twitter. 
First, the community feedback does not significantly affect most users 
(about 67\% and 85\% in Reddit and Twitter, respectively). 
Second, there are few users who are influenced negatively: only 0.2\% and 1\% of users in Reddit and Twitter, respectively. 
The corresponding analysis based on the number of up/down-votes on Reddit and the number of likes on Twitter gives quantitatively similar results as Table \ref{Tab:PosNeg_Authors} (Appendix A). 
As a sanity check, we measure whether including community feedback in the model improves predictive accuracy for the positive users, finding that it improves it by 0.19 $\pm$ 0.05~\% for Reddit and 0.53 $\pm$ 0.09~\% for Twitter. 
This result does not hold for users with insignificant susceptibility.


\begin{figure}[t]
 \centering
\includegraphics[width=8.5cm]{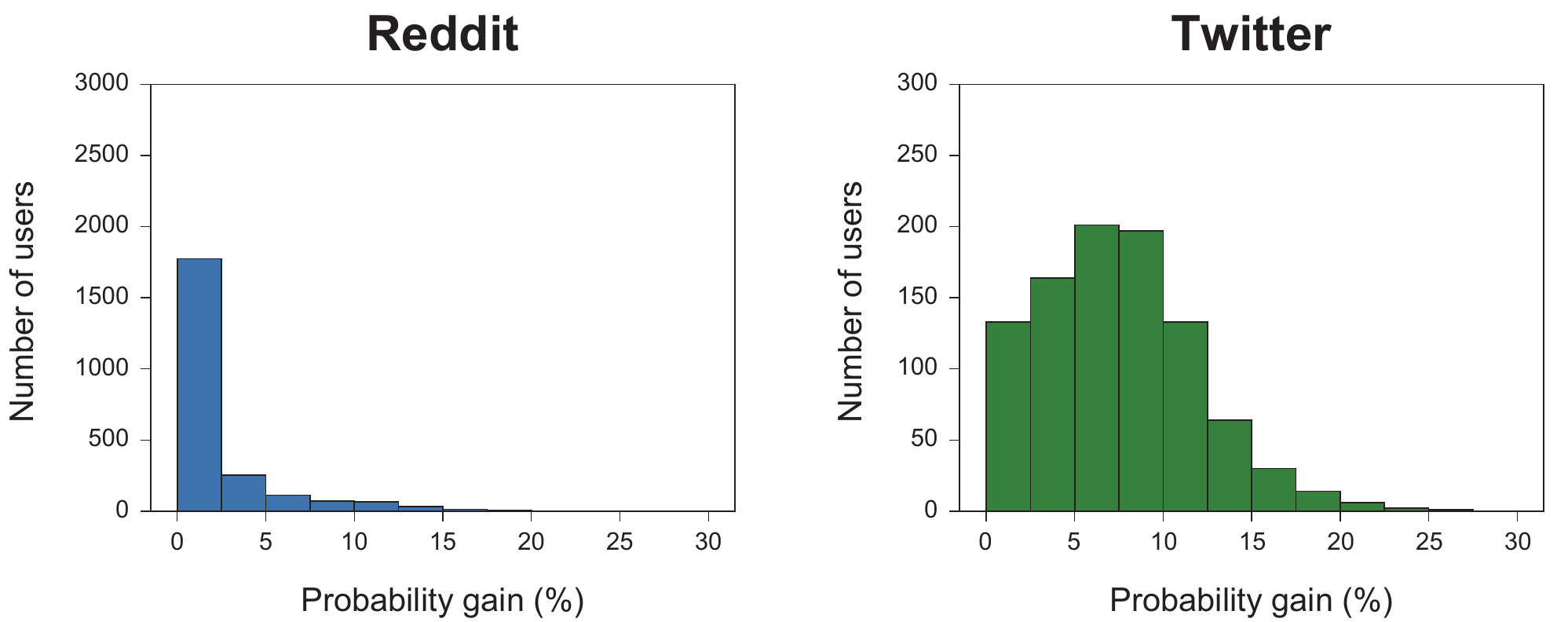}
    \caption{Gain of the topic repeat probability due to the community feedback $\Delta p_i$ for the positive users.}
    \label{Fig:Prob_Gain}
\end{figure} 

Finally, we evaluate the effect of the community feedback based on the probability gain of topic continuation due to reception of an extreme amount of feedback, for each user $i$:
\[ \Delta p_i= \frac{1}{N_i} \sum_{j=1}^{N_i} \left(P[ Y_{i, j}=1| f=0.99 ]- P[Y_{i, j}=1| f=0.5 ] \right), \]
where $N_i$ is the number of comments or tweets posted by the user, and $P[Y_{i,j}=1|f]$ is the probability of continuing the topic of the $j$-th post, while the feedback is fixed via an intervention to $f$~\cite{Pearl2009}.
The probability gain is averaged over user's posts, 
which estimates the expected increase in the probability of topic continuation due to reception of an extreme amount of feedback ($99\%$ percentile), in comparison to medium feedback ($50\%$ percentile). 
Figure~\ref{Fig:Prob_Gain} shows the distribution of the probability gain for the positive users. 
The community feedback alters slightly the probability of topic continuation: the median of the probability gain was 2\% (6\%) in Reddit (Twitter).
Finally, we calculate the effect size (Cohen's $d$) of the extreme $(f= 0.99)$ and median $(f=0.5)$ feedback.   
The effect size is 0.18 for Reddit and 0.29 for Twitter, indicating a small effect \cite{Cohen2013}.

\section{Discussion}
We investigate how community feedback affects individual users based on a predictive model. 
First, we have developed a model that predicts topic changes of an author by incorporating essential features: (i) author's properties, (ii) global topic trends due to news and social events, and (iii) the received feedback. 
Our model achieves high accuracy ($\approx$ 82\%) for two datasets from social media platforms (i.e., Reddit and Twitter). 
Then, we quantify the feedback effect on each user level using the model. While this effect does not significantly influence most users (67\% in Reddit and 85\% in Twitter), it affects the remaining users positively rather than negatively, i.e., these users are more inclined to continue the same topic if they receive positive feedback.

The effect of social feedback varies across different groups of users and social media platforms. 
The percentage of susceptible users is higher on Reddit than on Twitter, but the effect size is larger for the Twitter users than the Reddit users. 
We note also that in Reddit the percentage of susceptible users decreases with user activity, whereas it increases with user activity on Twitter (Tables~\ref{Tab:Activity_Dist_Reddit} and~\ref{Tab:Activity_Dist_Twitter} in Appendix F, respectively). 
Expert Twitter accounts often belong to celebrities or organizations, who may make use of social feedback in choosing their next topics to maximise engagement. 
This is not the case in Reddit, where user accounts have significantly lower visibility and organizations and celebrities do not have official accounts, hence there are less incentives for optimizing posting activity for engagement.
Future studies can test these hypotheses by distinguishing between different kinds of users (e.g., celebrities, organisations, casual users) in a given social media platform. 
Measuring the effect of community feedback for inactive users is more challenging, because they post less frequently.  
If users are extremely inactive but post in bursts, as it is often the case~\cite{Lehmann2012,Trilling2015}, the effect of community feedback can be captured by grouping similar users to obtain sufficient numbers of samples per each group.

At first glance the percentage of users susceptible to community feedback might appear to be small. However, Cheng et al. \cite{Cheng2014b} also report ``that negative feedback leads to significant behavioral changes that are detrimental to the community. [...] In contrast, positive feedback does not carry similar effects, and neither encourages rewarded authors to write more, nor improves the quality of their posts.'' While that study focused on other behavioral changes, repeating our setup while focusing on \emph{negative} feedback is a future direction to explore. 
Another reason that the percentage of susceptible users is small could be due to users getting accustomed to feedback and hence starting to ``price it in'' though certain expectations. For example, Cunha et al. \cite{Cunha2017} observe ``diminishing returns and social feedback on later posts is less important than for the first post.'' Though it is theoretically possible to look at changes in susceptibility over time, there are technical limitations related to obtaining complete user timelines. Still, differentiating between ``fresh'' and ``experienced'' users could be worth pursuing.


\subsection{Limitations:} 
This study has the following limitations. 
First, we focused on the number of comments (Reddit) and retweets (Twitter), but we did not consider the content or sentiment of the feedback. 
However, as discussed above, the effect of receiving negative feedback can be quite different from that of positive feedback~\cite{Cheng2014a}. 
While retweets typically imply positive feedback, such as support for the author and agreement with the tweet contents~\cite{Metaxas2015,Garimella2016}, comments and replies often contain a mixture or support and criticism~\cite{Garimella2016}. 
In our dataset, the positive, neutral, and negative comments accounted for about 40\%, 30\%, and 30\% for the total comments, respectively. This difference in sentiment is a possible reason why the effect of community feedback is smaller in Reddit than in Twitter. 
It would be interesting to extend the logistic model to incorporate the sentiment of the comments. At the same time, a negative sentiment does not necessarily indicate an antagonistic position towards the original post. For example, a post about a tragic event is likely to attract many comments with a negative sentiment, while agreeing with the original position. Stance detection \cite{10.1145/3369026} could hence be a useful direction to explore in the future.

Second, topic classification from short texts (e.g., tweet) is still a challenging task. While most of subreddit titles were interpretable for us, some topics extracted from tweets were not. This might be another reason why the results of Reddit and Twitter are different quantitatively (Tables \ref{Tab:PosNeg_Authors}). Note that noise in the topic classifier would lead to an underestimate of the effect that community feedback, or any other feature, has on topic continuation as the dependent variable, i.e. whether a topic is repeated or not, becomes more random and less predictable than it actually is. Hence, we believe that our estimates for the percentage of susceptible users and for the gains of the topic repeat probability due to community feedback are both lower bounds.

Third, we only looked at one type of behavior, topic continuation vs.\ topic change, and looked at effects averaged across all topics. Other behaviors, such as time until the next post or even churn probabilities could be looked at. Furthermore, the effect might be heterogeneous across topics. Future work is needed to look at different types of behavior change, as well as additional factors that might influence the effect heterogeneity.

Fourth, our current study does not look at who provides feedback, whether a close friend, an acquaintance, or a stranger. Previous work looking at fact-checking interventions for false statements on Twitter \cite{DBLP:conf/icwsm/HannakMKW14,doi:10.1080/10584609.2017.1334018} found that the type of social link did effect the likelihood to accept a fact-checking intervention. While the collection of social network information adds certain technical challenges related to API limits, the incorporation of such information seems a promising future direction.

Fifth, an additional, inherent challenge when collected data from online platforms is the fact that these platforms change for at least two reason: (i) Their user bases changes and, once no longer undergoing exponential growth, generally matures both in terms of expertise on the platform as well as in terms of biological age. (ii) Platforms periodically introduce new features, such as Twitter's ``retweet with comment" \cite{Garimella2016} or its expansion of the 140 character limit to 280 \cite{DBLP:conf/icwsm/GligoricA018}. In a sense, every new feature creates a new platform, making before-after generalizations difficult. While our method is expected to be applicable to future versions of the platforms studied, the quantitative findings might not be.

Sixth, our approach for estimating treatment effect based on predictive modeling may be affected by model misspecification.
We assume the logistic model and identify the confounding variables by exploring possible factors for the author's posting behavior. 
Although the high prediction accuracy (82\% for Reddit and Twitter) suggests that our predictive model is reasonable, there are many possible choices for the model and it is likely that more predictive models will be developed in future. 
For instance, it is interesting to extend the proposed model by incorporating the history of posting behavior of a user.
Additionally, similar to the matching methods, our method might miss confounding variables, which may affect the estimate of the community feedback effect. 
For example, our model neglects the temporal information, i.e., the time of previous posting. It would be interesting to develop such a predictive based on point process~\cite{Kobayashi2016}.
Our method can control some of unobserved confounders by including the global topic trend in the model. However, the topic trend is estimated by assuming that its effect on users is homogeneous, whereas the users may react differently to the topical events in reality.

Finally, it is possible that social feedback affects emotions more than observable actions such as topic choice. For example, Marayuma et al. \cite{marayumaetal17cscw0} observe that ``receiving positive feedback to social media posts instills a psychological sense of community in the poster.'' However, they do not report any actual behavior change. Reasoning about internal, mental states using social media is inherently challenging and something that this work does not attempt to do.




\subsection{Broader impact:} 

Our results contribute to the discussion on how operant conditioning affects social media users~\cite{Andreassen2015,Deibert2019} and suggest that social feedback systems are a critical and sensitive part of social media platforms that has an agenda-setting effect. 
The results of this study have implications for the design of social media. Prior studies show that social feedback influences opinions of consumers about online content and its propensity to spread~\cite{muchniketal13science,Grabowicz2015}, whereas this study shows its impact on authors' decisions on the topic to post next. 
We note that polarizing or biased topics receive more feedback than impartial topics~\cite{Zafar2016}. One can hypothesize that social influence contributes to this effect, by boosting the spread of topics that arouse emotions and elicit quick positive feedback from susceptible users. A potential solution addressing this issue is a novel design of social rating systems that accounts for susceptibilities of users.

Finally, we note that topic choice is a higher-level cognitive task~\cite{Maslow1943}, related to free will, so it is surprising that it is influenced by social feedback, although the father of operant conditioning considered free will an illusion~\cite{Skinner1938,Schacter2011}. It remains an open question how many of our choices are determined by various kinds of feedback, including social feedback, and how many are the result of free will.


\begin{backmatter}

\section*{Competing interests}
The authors declare that they have no competing interests.

\section*{Availability of data and materials}
The Reddit data used in this analysis can be freely downloaded from pushshift.io.\footnote{\url{https://files.pushshift.io/reddit/}}. The Twitter can also be crawled and used according to Twitter policy on data sharing
\section*{Author's contributions}
Grabowicz, Kobayashi and Weber conceptualized the study;  Kobayashi, Grabowicz, and Adelani devised the methodology; Grabowicz collected the data;  Adelani developed the software, performed the simulations and analyzed the data; Kobayashi, Grabowicz and Weber wrote the manuscript with input from all authors; 

\section*{Funding}
Kobayashi acknowledges support from the following sources: JSPS KAKENHI (grants JP17H03279, JP18K11560, and JP19H01133), JST ACT- I (grant JPMJPR16UC), and JST PRESTO (grant JPMJPR1925). Grabowicz acknowledges support from Volkswagen Foundation (grant 92136).

\end{backmatter}
\bibliographystyle{abbrv}
\bibliography{social_feedback}

\begin{thebibliography}{10}

\bibitem{Andreassen2015}
C.~S. Andreassen.
\newblock {Online Social Network Site Addiction: A Comprehensive Review}.
\newblock {\em Current Addiction Reports}, 2(2):175--184, 2015.

\bibitem{Aral2012}
S.~Aral and D.~Walker.
\newblock Identifying influential and susceptible members of social networks.
\newblock {\em Science}, 337:337--341, 2012.

\bibitem{Arganda2012}
S.~Arganda, A.~P{\'{e}}rez-Escudero, and G.~G. de~Polavieja.
\newblock {A common rule for decision making in animal collectives across
  species.}
\newblock {\em Proceedings of the National Academy of Sciences of the United
  States of America}, 109(50):20508--20513, dec 2012.

\bibitem{bernsteinetal13chi}
M.~S. Bernstein, E.~Bakshy, M.~Burke, and B.~Karrer.
\newblock Quantifying the invisible audience in social networks.
\newblock In {\em CHI}, pages 21--30, 2013.

\bibitem{Bhattacharya2020}
R.~Bhattacharya, R.~Nabi, and I.~Shpitser.
\newblock {Semiparametric Inference For Causal Effects In Graphical Models With
  Hidden Variables}.
\newblock {\em arXiv:2003.12659}, 2020.

\bibitem{Bishop2006}
C.~M. Bishop.
\newblock {\em Pattern recognition and machine learning}.
\newblock Springer-Verlag New York, 2006.

\bibitem{Boyd2004}
S.~Boyd, S.~P. Boyd, and L.~Vandenberghe.
\newblock {\em Convex optimization}.
\newblock Cambridge university press, 2004.

\bibitem{Cheng2014a}
J.~Cheng, L.~Adamic, P.~A. Dow, J.~M. Kleinberg, and J.~Leskovec.
\newblock Can cascades be predicted?
\newblock In {\em WWW}, pages 925--936, 2014.

\bibitem{Cheng2014b}
J.~Cheng, C.~Danescu-Niculescu-Mizil, and J.~Leskovec.
\newblock How community feedback shapes user behavior.
\newblock In {\em ICWSM}, pages 61--70, 2014.

\bibitem{Choromanska2015}
A.~Choromanska, M.~Henaff, M.~Mathieu, G.~B. Arous, and Y.~LeCun.
\newblock {The loss surfaces of multilayer networks}.
\newblock {\em J. Mach. Learn. Res.}, 38:192--204, 2015.

\bibitem{Cohen2013}
J.~Cohen.
\newblock {\em Statistical power analysis for the behavioral sciences}.
\newblock Academic press, 2013.

\bibitem{Conway2015}
B.~A. Conway, K.~Kenski, and D.~Wang.
\newblock {The Rise of Twitter in the Political Campaign: Searching for
  Intermedia Agenda-Setting Effects in the Presidential Primary}.
\newblock {\em Journal of Computer-Mediated Communication}, 20(4):363--380,
  2015.

\bibitem{Crane2008}
R.~Crane and D.~Sornette.
\newblock Robust dynamic classes revealed by measuring the response function of
  a social system.
\newblock {\em Proceedings of the National Academy of Sciences},
  105(41):15649--15653, 2008.

\bibitem{Cunha2017}
T.~Cunha, I.~Weber, and G.~Pappa.
\newblock A warm welcome matters!: The link between social feedback and weight
  loss in/r/loseit.
\newblock In {\em WWW}, pages 1063--1072, 2017.

\bibitem{Deibert2019}
R.~J. Deibert.
\newblock {The Road to Digital Unfreedom: Three Painful Truths About Social
  Media}.
\newblock {\em Journal of Democracy}, 30(1):25--39, 2019.

\bibitem{ecklesetal16pnas}
D.~Eckles, R.~F. Kizilcec, and E.~Bakshy.
\newblock Estimating peer effects in networks with peer encouragement designs.
\newblock {\em Proceedings of the National Academy of Sciences},
  113(27):7316--7322, 2016.

\bibitem{Forster1994}
M.~Forster and E.~Sober.
\newblock {How to Tell when Simpler, More Unified, or Less Ad Hoc Theories will
  Provide More Accurate Predictions}.
\newblock {\em British Journal of Philosophy of Science}, 45(January):1--35,
  1994.

\bibitem{frenchbazarova17jcmc}
M.~French and N.~N. Bazarova.
\newblock Is anybody out there?: Understanding masspersonal communication
  through expectations for response across social media platforms.
\newblock {\em Journal of Computer-Mediated Communication}, 22(6):303--319,
  2017.

\bibitem{Fujita2018}
K.~Fujita, A.~Medvedev, S.~Koyama, R.~Lambiotte, and S.~Shinomoto.
\newblock Identifying exogenous and endogenous activity in social media.
\newblock {\em Physical Review E}, 98(5):052304, 2018.

\bibitem{gaffneymatias18plos}
D.~Gaffney and J.~N. Matias.
\newblock Caveat emptor, computational social science: Large-scale missing data
  in a widely-published reddit corpus.
\newblock {\em PLOS ONE}, 13(7):1--13, 07 2018.

\bibitem{Garimella2016}
K.~Garimella, I.~Weber, and M.~De~Choudhury.
\newblock Quote rts on twitter: Usage of the new feature for political
  discourse.
\newblock In {\em WebSci}, pages 200--204, 2016.

\bibitem{Gelman2014}
A.~Gelman, J.~Hwang, and A.~Vehtari.
\newblock {Understanding predictive information criteria for Bayesian models}.
\newblock {\em Stat. Comput.}, 24(6):997--1016, 2014.

\bibitem{Ghosh2012}
S.~Ghosh, N.~Sharma, F.~Benevenuto, N.~Ganguly, and K.~Gummadi.
\newblock Cognos: Crowdsourcing search for topic experts in microblogs.
\newblock In {\em SIGIR}, pages 575--590, 2012.

\bibitem{DBLP:conf/icwsm/GligoricA018}
K.~Gligoric, A.~Anderson, and R.~West.
\newblock How constraints affect content: The case of twitter's switch from 140
  to 280 characters.
\newblock In {\em ICWSM}, pages 596--599, 2018.

\bibitem{przemyslaw}
P.~Grabowicz, M.~Babaei, J.~Kulshrestha, and I.~Weber.
\newblock The road to popularity: The dilution of growing audience on twitter.
\newblock In {\em ICWSM}, pages 567--570, 2016.

\bibitem{Grabowicz2015}
P.~A. Grabowicz, F.~Romero-Ferrero, T.~Lins, G.~G. de~Polavieja, F.~Benevenuto,
  and K.~P. Gummadi.
\newblock {An experimental study of opinion influenceability}.
\newblock {\em arXiv:1802.02163}, 2015.

\bibitem{Grinberg2016}
N.~Grinberg, P.~A. Dow, L.~A. Adamic, and M.~Naaman.
\newblock Changes in engagement before and after posting to facebook.
\newblock In {\em CHI}, pages 564--574, 2016.

\bibitem{grinbergetal17cscw}
N.~Grinberg, S.~Kalyanaraman, L.~A. Adamic, and M.~Naaman.
\newblock Understanding feedback expectations on facebook.
\newblock In {\em CSCW}, pages 726--739, 2017.

\bibitem{DBLP:conf/icwsm/HannakMKW14}
A.~Hannak, D.~Margolin, B.~Keegan, and I.~Weber.
\newblock Get back! you don't know me like that: The social mediation of fact
  checking interventions in twitter conversations.
\newblock In {\em ICWSM}, 2014.

\bibitem{Hirano2005}
K.~Hirano and G.~W. Imbens.
\newblock {The Propensity Score with Continuous Treatments}.
\newblock In {\em Appl. Bayesian Model. Causal Inference from Incomplete‐Data
  Perspect.}, chapter~7, pages 73--84. John Wiley {\&} Sons, Ltd, 2005.

\bibitem{Hoyer2009}
P.~O. Hoyer, D.~Janzing, J.~M. Mooij, J.~Peters, and B.~Sch{\"o}lkopf.
\newblock Nonlinear causal discovery with additive noise models.
\newblock In {\em NIPS}, pages 689--696, 2009.

\bibitem{Hutto2014}
C.~J. Hutto and E.~Gilbert.
\newblock Vader: A parsimonious rule-based model for sentiment analysis of
  social media text.
\newblock In {\em ICWSM}, 2014.

\bibitem{Imai2004}
K.~Imai and D.~A. {Van Dyk}.
\newblock {Causal inference with general treatment regimes: Generalizing the
  propensity score}.
\newblock {\em Journal of the American Statistical Association},
  99(467):854--866, 2004.

\bibitem{Kiciman2018}
E.~Kiciman, S.~Counts, and M.~Gasser.
\newblock Using longitudinal social media analysis to understand the effects of
  early college alcohol use.
\newblock In {\em ICWSM}, pages 171--180, 2018.

\bibitem{Kitagawa1996}
G.~Kitagawa and W.~Gersch.
\newblock {\em Smoothness priors analysis of time series}.
\newblock Springer Science \& Business Media, 1996.

\bibitem{Kobayashi2019}
R.~Kobayashi, S.~Kurita, A.~Kurth, K.~Kitano, K.~Mizuseki, M.~Diesmann, B.~J.
  Richmond, and S.~Shinomoto.
\newblock Reconstructing neuronal circuitry from parallel spike trains.
\newblock {\em Nature communications}, 10(1):1--13, 2019.

\bibitem{Kobayashi2016}
R.~Kobayashi and R.~Lambiotte.
\newblock Tideh: Time-dependent hawkes process for predicting retweet dynamics.
\newblock In {\em ICWSM}, pages 191--200, 2016.

\bibitem{Kobayashi2011}
R.~Kobayashi, Y.~Tsubo, P.~Lansky, and S.~Shinomoto.
\newblock Estimating time-varying input signals and ion channel states from a
  single voltage trace of a neuron.
\newblock In {\em NIPS}, pages 217--225, 2011.

\bibitem{Kramer2014}
A.~D.~I. Kramer, J.~E. Guillory, and J.~T. Hancock.
\newblock Experimental evidence of massive-scale emotional contagion through
  social networks.
\newblock {\em Proceedings of the National Academy of Sciences},
  111(24):8788--8790, 2014.

\bibitem{10.1145/3369026}
D.~K\"{u}\c{c}\"{u}k and F.~Can.
\newblock Stance detection: A survey.
\newblock {\em ACM Comput. Surv.}, 53(1), 2020.

\bibitem{latane}
B.~Latane.
\newblock {The psychology of social impact}.
\newblock {\em American Psychologist}, 36(4):343--356, Apr. 1981.

\bibitem{Latane1996}
B.~Latan{\'e} and T.~L'Herrou.
\newblock Spatial clustering in the conformity game: Dynamic social impact in
  electronic groups.
\newblock {\em Journal of personality and social psychology}, 70(6):1218, 1996.

\bibitem{Lee2011}
D.~Lee, H.~S. Kim, and J.~K. Kim.
\newblock The impact of online brand community type on consumer's community
  engagement behaviors: Consumer-created vs. marketer-created online brand
  community in online social-networking web sites.
\newblock {\em Cyberpsychology, Behavior, and Social Networking},
  14(1-2):59--63, 2011.

\bibitem{Lehmann2012}
J.~Lehmann, B.~Gon{\c{c}}alves, J.~J. Ramasco, and C.~Cattuto.
\newblock Dynamical classes of collective attention in twitter.
\newblock In {\em WWW}, pages 251--260, 2012.

\bibitem{Lipton2018}
Z.~C. Lipton.
\newblock {The mythos of model interpretability}.
\newblock {\em Commun. ACM}, 61(10):36--43, sep 2018.

\bibitem{litt12jbem}
E.~Litt.
\newblock Knock, knock. who's there? the imagined audience.
\newblock {\em Journal of Broadcasting \& Electronic Media}, 56(3):330--345,
  2012.

\bibitem{doi:10.1080/10584609.2017.1334018}
D.~B. Margolin, A.~Hannak, and I.~Weber.
\newblock Political fact-checking on twitter: When do corrections have an
  effect?
\newblock {\em Political Communication}, 35(2):196--219, 2018.

\bibitem{marayumaetal17cscw0}
M.~Maruyama, S.~P. Robertson, S.~Douglas, R.~Raine, and B.~Semaan.
\newblock Social watching a civic broadcast: Understanding the effects of
  positive feedback and other users’ opinions.
\newblock In {\em CSCW}, page 794–807, 2017.

\bibitem{Marwick2011}
A.~E. Marwick and D.~Boyd.
\newblock I tweet honestly, i tweet passionately: Twitter users, context
  collapse, and the imagined audience.
\newblock {\em New media \& society}, 13(1):114--133, 2011.

\bibitem{Maslow1943}
A.~H. Maslow.
\newblock {A theory of human motivation.}
\newblock {\em Psychological Review}, 50(4):370--396, 1943.

\bibitem{Matthews1975}
B.~W. Matthews.
\newblock Comparison of the predicted and observed secondary structure of t4
  phage lysozyme.
\newblock {\em Biochimica et Biophysica Acta (BBA)-Protein Structure},
  405(2):442--451, 1975.

\bibitem{McCombs1972}
M.~E. McCombs and D.~L. Shaw.
\newblock {The agenda-setting funcion of mass-media}.
\newblock {\em The Public Opinios Quarterly}, 36:176--187, 1972.

\bibitem{McCorriston2015}
J.~McCorriston, D.~Jurgens, and D.~Ruths.
\newblock Organizations are users too: Characterizing and detecting the
  presence of organizations on twitter.
\newblock In {\em ICWSM}, pages 650--653, 2015.

\bibitem{Metaxas2015}
P.~Metaxas, E.~Mustafaraj, K.~Wong, L.~Zeng, M.~O'Keefe, and S.~Finn.
\newblock What do retweets indicate? results from user survey and meta-review
  of research.
\newblock In {\em ICWSM}, pages 658--661, 2015.

\bibitem{muchniketal13science}
L.~Muchnik, S.~Aral, and S.~J. Taylor.
\newblock Social influence bias: A randomized experiment.
\newblock {\em Science}, 341(6146):647--651, 2013.

\bibitem{Naylor2012}
R.~W. Naylor, C.~P. Lamberton, and P.~M. West.
\newblock Beyond the “like” button: The impact of mere virtual presence on
  brand evaluations and purchase intentions in social media settings.
\newblock {\em Journal of Marketing}, 76(6):105--120, 2012.

\bibitem{Olteanu2017}
A.~Olteanu, O.~Varol, and E.~Kiciman.
\newblock Distilling the outcomes of personal experiences: A propensity-scored
  analysis of social media.
\newblock In {\em CSCW}, pages 370--386, 2017.

\bibitem{Pearl2009}
J.~Pearl.
\newblock {\em {Causality: Models, Reasoning and Inference}}.
\newblock Cambridge University Press, 2nd edition, 2009.

\bibitem{Perez2016}
R.~Perez-Vega, K.~Waite, and K.~O'Gorman.
\newblock Social impact theory: An examination of how immediacy operates as an
  influence upon social media interaction in facebook fan pages.
\newblock {\em The Marketing Review}, 16(3):299--321, 2016.

\bibitem{Rubin1974}
D.~B. Rubin.
\newblock Estimating causal effects of treatments in randomized and
  nonrandomized studies.
\newblock {\em Journal of educational Psychology}, 66(5):688--701, 1974.

\bibitem{Russell2010}
S.~Russell and P.~Norvig.
\newblock {\em {Artificial Intelligence: A Modern Approach}}.
\newblock Prentice Hall, imprint of Pearson Education, Inc., Upper Saddle
  River, New Jersey, 2010.

\bibitem{Schacter2011}
D.~L. Schacter, D.~T. Gilbert, and D.~M. Wegner.
\newblock {\em {Psychology (2nd Edition)}}.
\newblock Worth, New York, 2011.

\bibitem{scissorsetal16cscw}
L.~Scissors, M.~Burke, and S.~Wengrovitz.
\newblock What's in a like?: Attitudes and behaviors around receiving likes on
  facebook.
\newblock In {\em CSCW}, pages 1501--1510, 2016.

\bibitem{Shimizu2006}
S.~Shimizu, P.~O. Hoyer, A.~Hyv{\"a}rinen, and A.~Kerminen.
\newblock A linear non-gaussian acyclic model for causal discovery.
\newblock {\em Journal of Machine Learning Research}, 7(Oct):2003--2030, 2006.

\bibitem{Skinner1938}
B.~F. Skinner.
\newblock {\em {The behavior of organisms: an experimental analysis.}}
\newblock Appleton-Century, Oxford, England, 1938.

\bibitem{Spirtes2016}
P.~Spirtes and K.~Zhang.
\newblock Causal discovery and inference: concepts and recent methodological
  advances.
\newblock {\em Applied informatics}, 3(1):3, 2016.

\bibitem{Stuart2010}
E.~A. Stuart.
\newblock Matching methods for causal inference: A review and a look forward.
\newblock {\em Statistical science: a review journal of the Institute of
  Mathematical Statistics}, 25(1):1--21, 2010.

\bibitem{Trilling2015}
D.~Trilling.
\newblock {Two Different Debates? Investigating the Relationship Between a
  Political Debate on TV and Simultaneous Comments on Twitter}.
\newblock {\em Social Science Computer Review}, 33(3):259--276, 2015.

\bibitem{Xu2012}
Z.~Xu, Y.~Zhang, Y.~Wu, and Q.~Yang.
\newblock Modeling user posting behavior on social media.
\newblock In {\em SIGIR}, pages 545--554, 2012.

\bibitem{Yang2010}
Z.~Yang, J.~Guo, K.~Cai, J.~Tang, J.~Li, L.~Zhang, and Z.~Su.
\newblock Understanding retweeting behaviors in social networks.
\newblock In {\em CIKM}, pages 1633--1636, 2010.

\bibitem{Yin2014}
H.~Yin, B.~Cui, L.~Chen, Z.~Hu, and Z.~Huang.
\newblock A temporal context-aware model for user behavior modeling in social
  media systems.
\newblock In {\em SIGMOD}, pages 1543--1554, 2014.

\bibitem{Zafar2016}
M.~B. Zafar, K.~P. Gummadi, and C.~Danescu-Niculescu-Mizil.
\newblock Message impartiality in social media discussions.
\newblock In {\em ICWSM}, pages 466--475, 2016.

\bibitem{ZhangQ2015}
Q.~Zhang, Y.~Gong, Y.~Guo, and X.~Huang.
\newblock Retweet behavior prediction using hierarchical dirichlet process.
\newblock In {\em AAAI}, pages 403--409, 2015.

\bibitem{Zhao2011}
W.~X. Zhao, J.~Jiang, J.~Weng, J.~He, E.-P. Lim, H.~Yan, and X.~Li.
\newblock Comparing twitter and traditional media using topic models.
\newblock In {\em ECIR}, pages 338--349, 2011.

\end{thebibliography}


\renewcommand{\thetable}{A\arabic{table}}
\setcounter{table}{0}

\section*{Appendix }

\subsection*{A. Up/down-votes and likes as feedback}
In the main text, the feedback amount is calculated based on the number of comments or retweets. Since we have the timestamps for these individual pieces of feedback, we count only the feedback that was received before the next post is created, because only this feedback can causally influence the topic of the next post.
Here, we calculate the feedback amount based on the difference in the number of up-votes and down-votes on Reddit and the number of likes on Twitter. 
We examine the distribution of susceptibility among the users (Table \ref{Tab:PosNeg_Authors_fav}). 
The new results are quantitatively similar to the results of the main text (Table \ref{Tab:PosNeg_Authors}). 
It should be noted that the new results might be less causally valid since we do not have the timestamps of votes and likes. 

\vspace{0.5cm}
\begin{table}[h]
\centering
\footnotesize
    \begin{tabular}{lll}
    Group       		&  Reddit       		&  Twitter   \\ \hline
     \#Positive         &   2,450 (34.6\%)  	&    786 (11.4\%)  \\
     \#Negative        	&    96 (1.4\%)    	&      90(1.3\%)     \\
     \#Insignificant   	&   4,526 (64\%) 	&     6,006(87.3\%)   \\  \hline
    \end{tabular}     \vspace{0.5cm}
       \caption{Distribution of the susceptibility $\alpha_i$ to the community feedback (up/down-votes and likes).} 
    \label{Tab:PosNeg_Authors_fav}
\end{table}


\subsection*{B. Effect of topic granularity on Twitter results}
We investigate whether the number of topics, $K$, impacts our results and conclusions. %
First, we examine the effect of topic granularity on prediction accuracy.
When the number of topics is increased from $100$ to $200$, the accuracy slightly improves except for all the models: the accuracy is $70.56 \pm 0.00$\%, $74.90 \pm 0.01$\%, $80.82 \pm 0.00$\%, $81.21 \pm 0.00$\%, and $83.27 \pm 0.17$\% for the five models specified in Table~\ref{Tab:Pred_Accuracy}. 
Second, we examine the effect of topic granularity on the distribution of the susceptibility $\alpha_i$. With increased granularity the number of positive, negative, and insignificant users changes to $12\%$, $3\%$, and $85\%$, respectively (compare with Table~\ref{Tab:PosNeg_Authors}). 
Overall, the results do not change qualitatively when we change the topic granularity on Twitter. 
\newpage

\renewcommand{\thetable}{C\arabic{table}}
\setcounter{table}{0}

\subsection*{C. Detailed analysis of the prediction performance of the proposed model}
We evaluate the performance of the proposed model (Eq.~\ref{eq:user_model}) in predicting the topic change of a user. 
The proposed model considers three types of features: the user-dependent propensity to continue any topic (``Prop''), the preference to topics (``Pref''), and the topic trend due to news and social events (``Trend''). 
The prediction performance was evaluated by three metrics: 
the accuracy (``ACC''), the F1 score (``F1''), and the Matthews correlation coefficient (``MCC'')~\cite{Matthews1975,Kobayashi2019}, defined as 
\[
    ACC = \frac{TP+TN}{TP+TN+FP+FN}, 	
\]
\[
    F1 = \frac{2TP}{2TP+FP+FN}, 	
\]
\[
    MCC = \frac{TP \times TN- FP \times FN}{\sqrt{ (TP+FP) (TP+FN) (TN+FP) (TN+FN) } }, 
\]
where $TP$, $TN$, $FP$, and $FN$ is the number of true positives, true negatives, false positives, and false negatives, respectively, and the positive class corresponds to topic continuation. 
Tables~\ref{Tab:Pred_Full_Reddit} and~\ref{Tab:Pred_Full_Twitter} show the prediction performance for Reddit and Twitter, respectively. 
The results are qualitatively the same for the three measures. 
The topic preference of a user is a much more important feature than the propensity to topic continuation, and the topic trend further improves the prediction accuracy. 
\vspace{0.5cm}
\begin{table}[h]
\centering
\footnotesize
   \begin{tabular}{llll}
    Feature &  Accuracy &  F1   &  MCC \\ \hline
    None                   &  $63.43 \pm 0.01$\%   &  $38.81 \pm 0.00$\%   &   $0.00 \pm 0.00$\%   \\
    Prop                   &  $63.87 \pm 0.00$\%   &  $40.38 \pm 0.01$\%   &   $7.93 \pm 0.04$\%   \\ 
    Pref                   &  $79.34 \pm 0.01$\%   &  $77.36 \pm 0.00$\%   &   $54.85 \pm 0.00$\%  \\
    Prop, Pref             &  $79.30 \pm 0.01$\%   &  $77.24 \pm 0.00$\%   &   $54.67 \pm 0.01$\%  \\
    Prop, Pref, Trend      & $82.27 \pm 0.01\%$    &  $80.98 \pm 0.05$\%   &   $61.99 \pm 0.10$\% \\
    \hline
    \end{tabular} \vspace{0.5cm}
       \caption{Effect of author property and topic trend on the prediction performance (Reddit). 
       The mean and the 99\% confidence interval for 200 trials are shown. 
       }
    \label{Tab:Pred_Full_Reddit}
\end{table}

\begin{table}[b]
\centering
\footnotesize
   \begin{tabular}{llll}
    Feature &  Accuracy &  F1   &  MCC \\ \hline
    None                    &  $65.44 \pm 0.00$\%   &  $39.56 \pm 0.00$\%   &   $0.00\pm 0.00$\%   \\
    Prop                    &  $71.54 \pm 0.01$\%   &  $60.01 \pm 0.02$\%   &   $31.74 \pm 0.04$\%   \\ 
    Pref                    &  $80.29 \pm 0.01$\%   &  $77.02 \pm 0.00$\%   &   $54.98 \pm 0.00$\%  \\
    Prop, Pref              &  $80.52 \pm 0.01$\%   &  $77.53 \pm 0.00$\%   &   $55.68 \pm 0.01$\%  \\
    Prop, Pref, Trend       & $81.97 \pm 0.01\%$    &  $79.75 \pm 0.03$\%   &   $59.59 \pm 0.06$\% \\
    \hline
    \end{tabular} \vspace{0.5cm}
       \caption{Effect of author property and topic trend on the prediction performance (Twitter). 
       The mean and the 99\% confidence interval for 200 trials are shown.  
       }
    \label{Tab:Pred_Full_Twitter}
\end{table}
\newpage
\renewcommand{\thefigure}{D.\arabic{figure}}
\setcounter{figure}{0}

\subsection*{D. Similarity between the topic trend and the popularity of posts}
\begin{figure*}[!b]
 \centering
  \includegraphics[width=12cm]{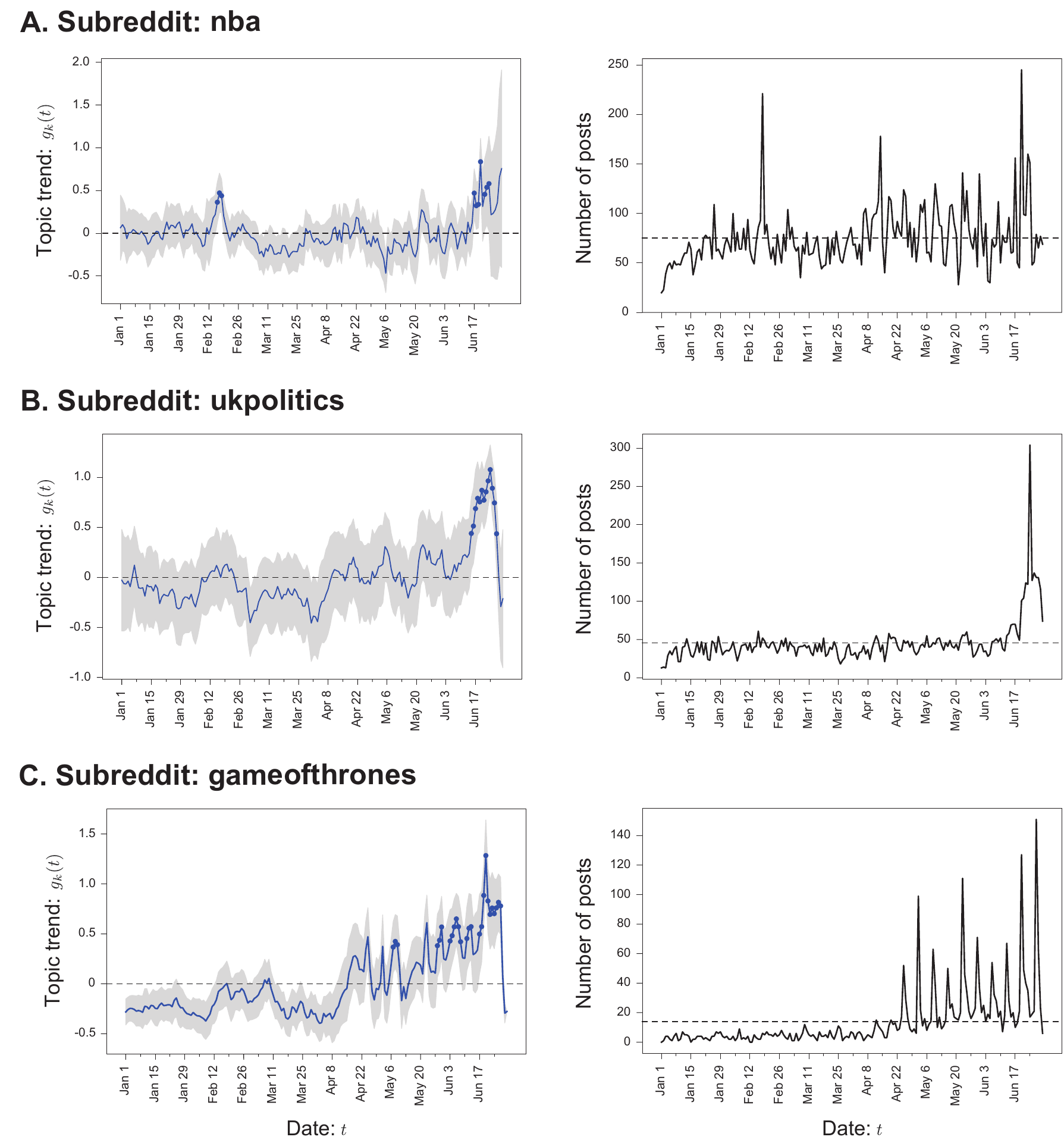}
	\caption{A comparison between the topic trend $g_k(t)$ (blue in the left panels) and the popularity of posts (black in the right panels) over time in Reddit. The gray area represents the 95\% confidence interval of the topic trend. Filled circles mark the days when the topic trend is significantly higher than zero for at least three consecutive days. 
	The three examples are the same as those in Figure~\ref{Fig:Topic_Trend}.}
    \label{Fig:Trend_Reddit}
\end{figure*}

\begin{figure*}[!t]
 \centering
  \includegraphics[width=12cm]{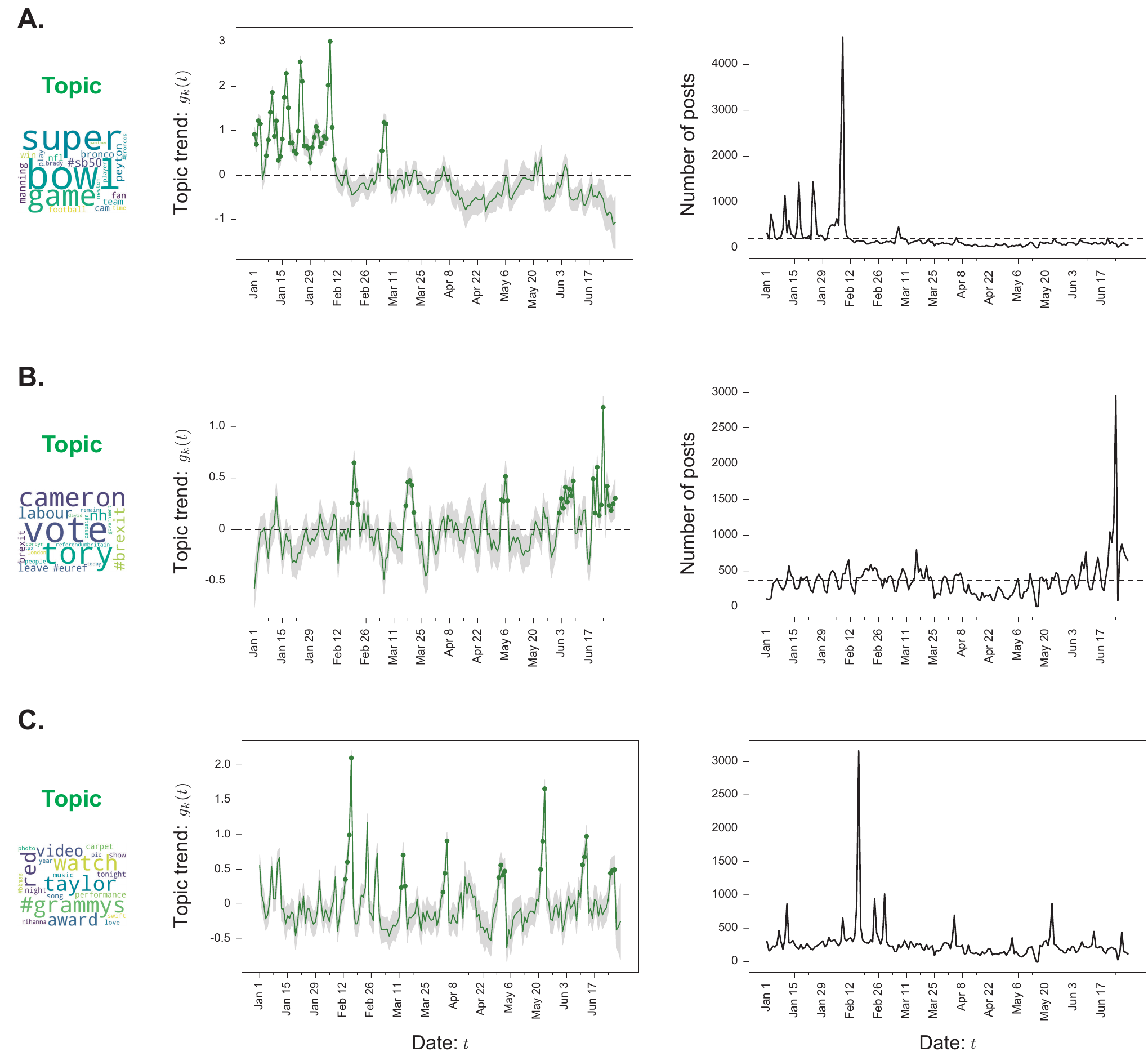}
	\caption{A comparison between the topic trend $g_k(t)$ (green in the left panels) and the popularity of posts (black in the right panels) over time in Twitter. The gray area represents the 95\% confidence interval of the topic trend. Filled circles mark the days when the topic trend is significantly higher than zero for at least three consecutive days. 
	The three examples are the same as those in Figure~\ref{Fig:Topic_Trend}.}
    \label{Fig:Trend_Twitter}
\end{figure*}

The topic trend $g_k(t)$ describes the effect of the topic on the probability of topic continuation in a subsequent post. Figures~\ref{Fig:Trend_Reddit} and~\ref{Fig:Trend_Twitter} show the comparison between the topic trend and the number of post per day on a given topic in Reddit and Twitter, respectively. 
Whereas the topic trend is similar to the popularity of the posts, they are different because the topic trend is defined as a logit of the probability and it is smoothed. 
%
%
%
\clearpage
\renewcommand{\thetable}{E\arabic{table}}
\setcounter{table}{0}
\subsection*{E. Comparison of the prediction performance among the feedback functions}
To determine the feedback function, we compare the prediction performance of three feedback functions $f$ (Eq. 1): (i) the feedback quantity $x$, (ii) its logarithm $\log(x)$, and (iii) its probability integral transform $P( X_i< x)$, 
where $P(X_i <x)$ is the probability that the given $i$-th user receives feedback smaller than $x$. 
We examined two feedback quantities $x$: (a) the feedback amount $n_i$, where $n_i$ is the number of comments (Reddit) or retweets (Twitter) from the previous post, and (b) the feedback rate $n_i/\Delta t$, where $\Delta t$ is the duration from the previous post. 
Table~\ref{Tab:Pred_FAmount} and \ref{Tab:Pred_FRate} show the prediction accuracy based on the feedback amount and rate, respectively. 
Feedback rate improves the prediction accuracy in comparison to feedback amount. 
We adopt the cumulative probability of the feedback rate, $P(R_i< r_i)$, as the feedback function, because it achieves the best accuracy. 
\vspace{0.5cm}
\begin{table}[htb]
  \caption{The effect of feedback function on the prediction accuracy (the feedback amount $x= n_i$). }
  \vspace{0.5cm}
  	\begin{tabular}{l|cc}
     	Feedback function	&	Reddit					&	Twitter	 	\\		\hline
     	None                &   $82.27 \pm 0.04 $\%     &   $81.97 \pm 0.02 $\%     \\
		$n_i$			    & $81.57 \pm  0.03 $\%	& 	$81.80 \pm  0.03 $\%		\\
		$\log(n_i)$		    &	$81.82 \pm  0.04$\%     & 	$81.91 \pm  0.03$\%	    \\
	   	$P(N_i< n_i)$	    &	$82.02 \pm  0.05 $\%	&	$81.95 \pm  0.03 $\%
  \end{tabular}     \label{Tab:Pred_FAmount}
\end{table}
\vspace{0.5cm}
\begin{table}[htb]
  \caption{The effect of feedback function on the prediction accuracy (the feedback rate $x= r_i$). }
  \vspace{0.5cm}
  	\begin{tabular}{l|cc}
     	Feedback function	&	Reddit				    &	Twitter		\\		\hline
     	None                &   $82.27 \pm 0.04 $\%     &   $81.97 \pm 0.02 $\%     \\
		$r_i$			    &	$82.35 \pm  0.03 $\%	&	$81.99 \pm  0.03 $\%		\\
		$\log(r_i)$		    &	$82.19 \pm  0.03 $\% 	& 	$ 82.00 \pm 0.02$\%	\\
	   	$P(R_i< r_i)$   	&	$82.41 \pm  0.04 $\%	&	$82.07 \pm  0.02$\%
  \end{tabular}      \label{Tab:Pred_FRate}
\end{table}

\clearpage
\renewcommand{\thetable}{F\arabic{table}}
\setcounter{table}{0}
\subsection*{F. Dependency of the distribution of susceptibility on posting activity}
In the main text, we focused on the expert users who post actively, that is, they post more than $50$ posts in six months. 
We examine the dependency of the distribution of the susceptibility (Table~\ref{Tab:PosNeg_Authors}) on the posting rate. 
The users in Reddit and Twitter were divided into four groups (G1, G2, G3, and G4) based on the number of posts, each group having the same sample size. 
The number of posts by the users in G1 ranges from 50 to $Q1$, where $Q1$ is the first quartile of the number of posts. The number of posts by the users in G2 ranges from $Q1$ to $Q2$, where $Q2$ is the second quartile of the number of posts. In the same way, the number of posts by the users in G3 ranges from $Q2$ to $Q3$, and the number of posts by the users in G4 is above $Q3$. The quartiles of the number of posts were $71$, $92$, and $139$ in Reddit, and $133$, $281$, and $666$ in Twitter. 
Then, we compared the susceptibility distribution among the four groups. In Reddit, the fraction of users who are positively influenced drops with increasing posting rate (Table~\ref{Tab:Activity_Dist_Reddit}), suggesting that more active users tend to be less susceptible. In Twitter, this fraction increases with the posting rate (Table~\ref{Tab:Activity_Dist_Twitter}). 
\vspace{0.5cm}

\begin{table}[hb]
\centering
    \begin{tabular}{llll}
           &  \#Positive    & \#Negative  &  \#Insignificant  \\ \hline
    G1     &  761 (42\%)  &  1 (0\%)    &   6,141 (58\%)  \\
    G2     &  577 (33\%)  &  1 (0\%)    &   1166 (67\%)  \\
    G3     &  509 (29\%)  &  0 (0\%)    &   1,516 (71\%)  \\ 
    G4     &   470 (27\%)   &  11 (1\%)    &  1,268 (72\%)  \\ \hline
    \end{tabular}
       \caption{Dependence of the susceptibility distribution on the posting rate for Reddit. The users are divided into four equal-sized groups (G1, G2, G3, and G4) based on their posting rate.}
    \label{Tab:Activity_Dist_Reddit}
\end{table}

\begin{table}[hb]
\centering
    \begin{tabular}{llll}
           &  \#Positive    & \#Negative  &  \#Insignificant  \\ \hline
    G1     &    104 (6\%)   &  28 (2\%)    &  1,583 (92\%)     \\
    G2     &    119 (7\%)   &  9 (1\%)    &  1,581 (92\%)  \\
    G3     &    236 (14\%)   &  14 (1\%)    &  1,467 (85\%)  \\
    G4     &    486 (28\%)   &  34 (2\%)    &  1,199 (70\%)  \\
    \hline
    \end{tabular}
       \caption{Dependency of the susceptibility distribution on the posting rate for Twitter.}
    \label{Tab:Activity_Dist_Twitter}
\end{table}
\end{document}